\documentclass[amsmath,amssymb,aps,prd,reprint,groupedaddress,showpacs]{revtex4}

\usepackage{graphicx}
\newcommand{\im}{{i}} % may be \rm i or something else
\newcommand{\pd}{\partial}
\newcommand{\g}{\gamma}

\newcommand{\bs}{\boldsymbol}

\providecommand{\cal}{\mathcal}
\renewcommand{\ol}{\overline} % we don't need \ol from revtex
\newcommand{\ti}{\tilde}

\newcommand{\sgn}{\mathop{\rm sgn}\nolimits}

\newcommand{\Tr}{\mathop{\rm Tr}\nolimits}
\newcommand{\Det}{\mathop{\rm Det}\nolimits}

\newcommand{\rarr}{\rightarrow}

\newcommand{\la}{\langle}
\newcommand{\ra}{\rangle}

\newcommand{\HD}{H_{\rm D}}

\begin{document}

\title{Chiral density waves in quark matter within the Nambu--Jona-Lasinio model in an external magnetic field}

\author{I.~E.~Frolov}
\email[]{igor.e.frolov@gmail.com}
\author{V.~Ch.~Zhukovsky}
\email[]{zhukovsk@phys.msu.ru}
\affiliation{Department of Theoretical Physics, Faculty of Physics, Moscow State University, 119991, Moscow, Russia}

\author{K.~G.~Klimenko}
\email[]{kklim@ihep.ru}
\affiliation{Institute for High Energy Physics, 142281, Protvino, Moscow Region, Russia}
\affiliation{Dubna University, Protvino Branch, 142281, Protvino, Moscow Region, Russia}

\date{\today}

\begin{abstract}
A possibility of formation of static dual scalar and pseudoscalar
density wave condensates in dense quark matter is considered for the
Nambu--Jona-Lasinio model in an external magnetic field. Within a
mean-field approximation, the effective potential of the theory is
obtained and its minima are numerically studied; a phase diagram of
the system is constructed. It is shown that the presence of a magnetic
field favors the formation of spatially inhomogeneous 
condensate configurations at low temperatures and
arbitrary nonzero values of the chemical potential.
\end{abstract}


\maketitle

% ======================================

\section{Introduction\label{sec:intro}}

At present, one of the most commonly used effective theories of
quantum chromodynamics is the Nambu--Jona-Lasinio (NJL) model
\cite{bib:NJL1, bib:NJL2}, a local relativistic four-fermion
interaction theory. The QCD and NJL Lagrangians possess the same
symmetry group, the NJL model is therefore widely exploited in
studying the nonperturbative QCD vacuum and its properties under
various external conditions. Many features of quarks and light mesons
can be successfully described within the NJL model on the basis of
the spontaneous chiral symmetry breaking phenomenon
\cite{bib:EbR, bib:Klevansky_NJL, bib:EbRV}.

Considering the QCD ground state properties, a number of studies were
dedicated to the possibility of formation of spatially nonuniform
phases in dense quark matter. It was first shown~\cite{bib:DGR} that
spatially inhomogeneous and anisotropic chiral condensation may occur
in QCD at asymptotically high values of the chemical potential and
large $ N_{c} $, the ground state spatial structure taking the form of
a standing wave. This phenomenon was further discussed in literature
\cite{bib:qcd_cond_wave_1, bib:qcd_cond_wave_2, bib:qcd_cond_wave_3}
investigating the possibility of such type of symmetry breaking and
its competition with color superconductivity under various conditions
including intermediate densities of quark matter. 
The problem has also been examined recently in the context of quarkyonic matter \cite{bib:Kojo}.
Along with QCD
studies, similar behavior of the ground state has also been
discovered and successfully reproduced in NJL-like effective models
\cite{bib:qcd_cond_wave_3, bib:Dautry_Nyman, bib:Tatsumi_old, bib:Broniowski_Kutschera_1,
bib:KBK2, bib:Tatsumi_new1, bib:Tatsumi_new2, bib:Ohwa, bib:Thies,
bib:Sadzikowski_Broniowski, bib:Nakano_Tatsumi_dcdw,
bib:Sadzikowski_2, bib:Partyka_Sadzikowski,
bib:Sedrakian_Rischke, bib:BFKT, bib:Maedan, bib:Partyka_1, bib:Polyakov_loop_new}, although
the effect tends to be dependent on the adopted regularization scheme
(see, e.g., Ref.~\cite{bib:Broniowski_Kutschera_amb} for details).

Spatially nonuniform condensates proposed at the start
of the theoretical research on the subject
\cite{bib:Dautry_Nyman} and studied extensively later on are known as
dual chiral density waves (DCDW, the name introduced in
Ref.~\cite{bib:Nakano_Tatsumi_dcdw}). 
The corresponding configuration can be described as follows: 
\begin{equation}
\label{eq:cond_def}
	\begin{aligned}
	    \la \ol \psi \psi \ra                     &= \Delta \cos {\bf qr}, \\
	    \la \ol \psi \im \g^{5} \tau_{3} \psi \ra &= \Delta \sin {\bf qr}, \\
	\end{aligned}
\end{equation}
where $ \Delta $ is the chiral density amplitude, $ \bf q $ is a
wave vector (which has to be determined dynamically along with $
\Delta $), and $ \tau_{a} $ are the isospin Pauli matrices. 
Expectation values $ \la \ol \psi \psi \ra $ and $ \la \ol
\psi \im \g^{5} \tau_{3} \psi \ra $ are identified with $ \sigma $ and
$ \pi^{0} $ condensates; one generally assumes $ \la \ol \psi \im
\g^{5} \tau_{1} \psi \ra = \la \ol \psi \im \g^{5} \tau_{2} \psi \ra =
0 $, thus charged $ \pi^{\pm} $ condensates being absent. In general,
scalar and pseudoscalar condensates are on the chiral circle: $ \la
\ol \psi \psi \ra^{2} + \la \ol \psi \im \g^{5} {\bs \tau} \psi \ra^{2}
= \Delta^{2} $. It is argued (see, e.g.,
Refs.~\cite{bib:Sadzikowski_Broniowski, bib:Nakano_Tatsumi_dcdw}) that
DCDW may arise between the massive and symmetric phases of the NJL
model at low temperatures if the coupling constant is sufficiently
large. The formation of DCDW along with color superconductivity
has also been discussed in literature
\cite{bib:Sadzikowski_2, bib:Partyka_Sadzikowski, bib:Sedrakian_Rischke}.
It should be noted however that,
although the majority of studies of condensate inhomogeneity focus on wavelike configurations and DCDW in particular,
this is mainly for technical reasons.
There may exist other competing and even more preferable spatially nonuniform ground state configurations like domain walls,
see, e.g., Ref. \cite{bib:Nickel},
but, in general, they are much harder to deal with.
For the same technical reasons, one usually considers 
the limit of vanishing quark current masses (chiral limit), 
although recently efforts to get rid of this assumption have been made 
\cite{bib:BFKT, bib:Maedan, bib:Partyka_1}.

The chiral condensation phenomenon (with spatially homogeneous condensate configurations)
has recently attracted great
attention in the situation when external gauge fields
and, in particular, strong magnetic fields are present
\cite{bib:HCMF1, bib:HCMF2, bib:HCMF3, bib:HCMF4, bib:HCMF5,
bib:HCMF6, bib:HCMF7, bib:HCMF8, bib:HCMF9, bib:HCMF10, bib:HCMF11}.
In fact these fields are common in the
physical circumstances where the phase structure of quark matter is of
interest, e.g., in compact stars or in heavy-ion collision processes
\cite{bib:heavy_ion_collisions}.
The effect of magnetic fields on the spatially nonuniform
chiral condensation is therefore worth investigation.
This is important in the
context of a targeted experimental search for possible condensate
inhomogeneity signals predicted, e.g., in
Ref.~\cite{bib:Sadzikowski_Broniowski}. Some interesting results
related to the subject have been obtained in
Ref.~\cite{bib:Son_Stephanov} showing that a stack of $ \pi^{0} $
domain walls may emerge in the QCD vacuum in a strong magnetic field
due to the axial anomaly (and concerning color superconductivity in a
magnetic field, see, e.g., Refs.~\cite{bib:CSCMF_2, bib:Noronha_Shovkovy}
suggesting
the formation of magnetic domains in cores of compact stars). 

In this paper, we examine the formation of the condensate configuration
defined in Eq. (\ref{eq:cond_def}) in dense quark matter in the framework of
the NJL model in the presence of an external magnetic field and show that
the latter favors the emergence of DCDW at low temperatures. 
Limiting ourselves to the chiral limit,
we base our calculations upon exact solutions of the Dirac equation and use the
proper-time regularization method 
such that our results agree with
Ref.~\cite{bib:Nakano_Tatsumi_dcdw} in the zero-field limit.

% ======================================

\section{The Model\label{sec:model}}

We start from the NJL Lagrangian density for a quark field $ \psi $
with $ N_{f} = 2 $ flavors (representing the up- and down-quarks) and $
N_{c} = 3 $ colors: 
\begin{equation}
\label{eq:L_NJL}
    {\cal L}^{\rm NJL}
    = \ol \psi \left(\im \g D - m_{c} + \mu \g^{0} \right) \psi
    + G \left[ (\ol \psi \psi)^{2} + (\ol \psi \im \g^{5} \bs \tau \psi)^{2} \right],
\end{equation}
where $ G $ is the coupling constant, $ \mu $ is the chemical
potential, $ m_{c} $ is the quark current mass, the covariant
derivative $ D = \pd + \im Q A $ with $ A $ being the electromagnetic
field, and $ Q $ the electric charge matrix acting in the flavor space: 
\begin{equation*}
	Q =
	\begin{pmatrix}
		\frac{2}{3} e & 0 \\
		0 & -\frac{1}{3} e \\
	\end{pmatrix}, \quad
	e > 0.
\end{equation*}
We take $ \g^{5} = -\im \g^{0} \g^{1} \g^{2} \g^{3} $ and assume $
\tau_{3} $ to be diagonal, we use the standard (Dirac) representation
of the $ \g $ matrices throughout the paper; matrix indices are
suppressed in our notation when possible. In what follows, we hold $
m_{c} = 0 $ assuming that the appropriate dimensional parameters in
our model tend to be much greater than $ m_{c} \simeq 5 \, \text{MeV} $. 
The symmetry of the model is therefore $ SU_{L}(2) \times SU_{R}(2) $
and it is reduced to $ U_{\tau_{3}L}(1) \times U_{\tau_{3}R}(1) $
when an external homogeneous magnetic field is present
(with the field strength pointing in the $ z $ direction).

Using ansatz (\ref{eq:cond_def}), we obtain the Lagrangian density in
the mean-field approximation (we only take into account Hartree terms
here, see a discussion on this subject in
Ref.~\cite{bib:Nakano_Tatsumi_dcdw}): 
\begin{equation}
\label{eq:L_MF}
    {\cal L}^{\rm MF}
    = \ol \psi \left[
        \im \g D + \mu \g^{0}
        - m \left( \cos {\bf qr} + \im \g^{5} \tau_{3} \sin {\bf qr} \right)
    \right] \psi
    - \frac{m^{2}}{4G},
\end{equation}
where we have denoted $ m = -2 G \Delta $. We assume that the system
resides in an external magnetic field, the 
wave vector $ \bf q $
being parallel to the field strength $ \bf H $, both vectors oriented
along the $ z $ axis. Such an assumption is reasonable due to the
symmetry considerations; possible small deviations of $ \bf q $
from the preferred orientation along $ \bf H $
are taken into account further.

As it is commonly done when considering model (\ref{eq:L_MF}), we use
a field transformation $ \psi \rarr e^{\im \g^{5} \tau_{3} bx } \, \psi
$, $ \ol \psi \rarr \ol \psi \, e^{\im \g^{5} \tau_{3} bx } $, where $
b^{\mu} \equiv (0, {\bf b}) $, $ x^{\mu} \equiv (t, {\bf r}) $, and $
{\bf b} = {\bf q} / 2 $, to remove the spatial modulation from the
resulting Lagrangian density~$ \cal L $: 
\begin{equation}
\label{eq:L}
  	{\cal L}
	= \ol \psi \left(
   		\im \g D + \mu \g^{0} - m + \g^{5} \tau_{3} \g b
    \right) \psi
    - \frac{m^{2}}{4G}.
\end{equation}
It should be noted, however, that special care is needed when
performing such operations in the presence of background gauge
fields. To obtain correct results, one should apply, for example, Fujikawa's
method~\cite{bib:Fujikawa_method} and its generalizations for finite
fermion field transformations. Fortunately, the path integral measure
$ {\cal D} \ol \psi {\cal D} \psi $ remains invariant in our case
since the quantity $ \epsilon^{\mu \nu \alpha \beta} F_{\mu \nu}
F_{\alpha \beta} $ which arises in Fujikawa's exponent, where $ F $ is
the electromagnetic field strength and $ \epsilon $ is the
antisymmetric tensor, vanishes in the absence of an electric field. 

In what follows, we obtain the thermodynamic potential $ \Omega $ for
the model described by Eq. (\ref{eq:L}) and then study numerically 
the minima of $ \Omega $ with respect to the order parameters $ m $ and $ {\bf b} $. 

% ======================================

\section{One-Particle Energy Spectrum\label{sec:spec}}

For later convenience, let us first 
consider a simplified model for a charged fermion (electron)
field having no flavors or colors with the Lagrangian density
\begin{equation}
\label{eq:L_electron}
  	{\cal L}
	= \ol \psi \left(
   		\im \g D - m - \g^{5} \g b
    \right) \psi,
\end{equation}
where $ D = \pd - \im e A $, $ e > 0 $. The term $ \ol \psi \g^{5} \g b
\psi $ in the latter expression describes a Lorentz- and CPT-breaking
background interaction controlled by the axial four-vector $ b^{\mu}
$. This type of interaction arising within the context of the Standard
Model Extension~\cite{bib:Colladay_Kostelecky} has been a subject of
extensive theoretical research in recent years (see, e.g., Refs.
\cite{bib:EZR, bib:ZLM, bib:Kharlanov_Zhukovsky_1, bib:Frolov_Zhukovsky}).
In this paper, in order to obtain
the one-particle energy spectrum of model (\ref{eq:L_electron}), 
we use a technique similar to that adopted in Ref.~\cite{bib:Frolov_Zhukovsky}. 

Let $ {\bf b} = (0, 0, b) $, $ {\bf H} = (0, 0, H) $, $ H > 0 $; we
take the electromagnetic field in the Landau gauge: $ A^{\mu} = (0,
{\bf A}) $, $ {\bf A} = (0, H x, 0) $. The modified Dirac Hamiltonian
derived from Eq. (\ref{eq:L_electron}) is as follows: 
\begin{equation}
\label{eq:H_electron}
	\HD = {\bs \alpha \bf P} + \g^{0} m - \Sigma_{3} b,
\end{equation}
where $ {\bf P} = -\im \nabla + e {\bf A} $ is the gauge-invariant
kinetic momentum, $ \bs \alpha = \g^{0} \bs \g $, $ \Sigma_{i} =
\frac{1}{2} \epsilon_{ijk} \sigma^{jk} $, $ \sigma^{\mu \nu} =
\frac{\im}{2} [ \g^{\mu}, \gamma^{\nu} ] $. Since $ [({\bs \alpha} {\bf
P}_{\bot})^{2}, \HD] = 0 $ where $ {\bf P}_{\bot} = (P_{1}, P_{2},
0) $, $ ({\bs \alpha} {\bf P}_{\bot})^{2} $ being an observable with
an oscillatorlike spectrum, it is easy to prove that the
eigenfunctions of $ \HD $ have a standard general form (see Chapter IV
of Ref.~\cite{bib:book_rel_el} for details): 
\begin{equation}
\label{eq:Psi}
    \Psi_{nqp}(x,y,z) =
    \frac{1}{\sqrt{2\pi}} \, e^{\im p z}
    \frac{1}{\sqrt{2\pi}} \, e^{\im q y}
    {
	    \left(
    	\setlength\arraycolsep{0.5pt}
    	\begin{array}{rl}
        	c_{1}     & u_{n-1}(\xi) \\
        	\im c_{2} & u_{n  }(\xi) \\
        	c_{3}     & u_{n-1}(\xi) \\
        	\im c_{4} & u_{n  }(\xi) \\
    	\end{array}
    	\right)
    }
    (eH)^{1/4}, \quad
    \xi = \sqrt{eH} x + \frac{q}{\sqrt{eH}},
\end{equation}
where $ u_{n}(\xi) $ are the orthonormalized Hermite functions [we
assume $ u_{-1}(\xi) \equiv 0 $] and $ \{ c_{i} \} $ are
spin-dependent coefficients. The quantum number $ n = 0, 1, \ldots $
is the Landau level, $ p $ is the momentum component parallel to the
magnetic field direction, and $ q $ is related to the symmetry center
$ x_{0} $ of the wavefunction $ \Psi $ along the $ x $ axis: $ q =
-x_{0} \, eH $. For each $ n > 0 $ and fixed $ q $ and $ p $, we have
an eigenvalue problem for a $ 4 \times 4 $-sized matrix $ K $ acting
on the vector $ \{ c_{i} \} $, where 
\begin{equation}
\label{eq:K}
	K = \alpha_{1} p_{\bot} + \alpha_{3} p + \g^{0} m - \Sigma_{3} b, \quad p_{\bot} = \sqrt{2eHn}.
\end{equation}
The quantity $ q $ is absent in Eq. (\ref{eq:K}) thus providing the
degeneracy of the energy spectrum with respect to it; this phenomenon
is related to the freedom in placing the particle's orbit in a
magnetic field and is preserved for any gauge of~$ {\bf A} $. 

Let us now consider a unitary transformation: $ \ti{K} = U^{-1} K U $
where $ U = e^{\im \Sigma_{2} \frac{\pi}{2}} e^{\im \g^{0} \Sigma_{2}
\frac{\pi}{2}} = \frac{1}{2} (1 + \im \Sigma_{2}) (1 + \im \g^{0}
\Sigma_{2}) $; it yields
\begin{equation}
\label{eq:K_tilde}
	\ti{K} = \alpha_{1} \ti{p}_{\bot} + \alpha_{3} \ti{p} + \g^{0} m + \g^{0} \Sigma_{3} \ti{\mu} H,
	\quad \ti{p}_{\bot} = p, \quad \ti{p} = -p_{\bot}, \quad \ti{\mu} H = b.
\end{equation}
The matrix $ \ti{K} $ formally corresponds to an electron with an
effective vacuum magnetic moment moving in an effective external
magnetic field. The problem for this case has been studied and solved
in Ref.~\cite{bib:Ternov_Bagrov_Zhukovsky} (note that the form of the
coefficients $ \{ c_{i} \} $ is independent of the adopted
electromagnetic field gauge). The case $ n = 0 $ requires a separate
treatment though, since $ K $ is reduced to a $ 2 \times 2 $-sized
matrix $ K_{0} $ acting on the coefficients $ \{ c_{i} \} $, $ i = 2,
4 $: 
\begin{equation*}
	K_{0} =
	\begin{pmatrix}
		m + b  & -p \\
		-p & -m + b \\
	\end{pmatrix}.
\end{equation*}
The eigenvalue problem for $ K_{0} $ can easily be solved. The final
expression for the energy spectrum has the form
\begin{equation}
\label{eq:E}
    E_{np\zeta\epsilon} =
    \begin{cases}
        \epsilon \sqrt{ \left( \zeta \sqrt{m^{2} + p^{2}} + b \right)^{2} + 2eHn}, & n = 1, 2, \ldots, \\[6pt]
        \epsilon \sqrt{m^{2} + p^{2}} + b, & n = 0, \\
    \end{cases}
\end{equation}
where $ \zeta = \pm 1 $ is the spin quantum number, $ \epsilon = \pm 1
$ is the energy sign (when $ n > 0 $). When $ n = 0 $, one only has
two (instead of four for $ n > 0 $) energy branches distinguished by
the number $ \epsilon $ and the latter has lost its meaning of the
energy sign in the presence of $ b \neq 0 $. Spectrum (\ref{eq:E}) has
been known in literature~\cite{bib:Miransky_spectrum} but the energy
shift of the $ n = 0 $ level has not been shown explicitly in the
paper cited. The specific asymmetry between the particle and
antiparticle energy spectra is due to the CPT-odd nature of the
background interaction present in our model. The phenomenon does not
manifest itself for free particles since one can compensate the
CPT-induced transformation $ {\bf b} \rarr -{\bf b} $ by a spatial
rotation. But this can no longer be done in the presence of a preferred
spatial direction which is introduced with $ {\bf H} $ in our
problem. 

The coefficients $ \{ c_{i} \} $ which meet the orthonormalization
requirement for the eigenfunctions $ \{ \Psi_{nqp\zeta\epsilon} \} $
are as follows: 
\begin{equation}
\label{eq:c_vector}
    \begin{pmatrix}
        c_{1} \\
        c_{2} \\
        c_{3} \\
        c_{4} \\
    \end{pmatrix}
    = \frac{1}{2 \sqrt{2}}
    \left(
    \setlength\arraycolsep{0.5pt}
    \begin{array}{r}
       	-\zeta B ( P - \epsilon \zeta Q ) \\
       	-A ( P + \epsilon \zeta Q )       \\
       	A  ( P - \epsilon \zeta Q )       \\
       	-\zeta B ( P + \epsilon \zeta Q ) \\
    \end{array}
    \right),
\end{equation}
where
\begin{equation*}
    A = \sqrt{1 + \frac{m}{\Pi}}, \quad
    B = \sqrt{1 - \frac{m}{\Pi}}, \quad
    P = \sqrt{1 - \frac{p_{\bot}}{E}}, \quad
    Q = \sqrt{1 + \frac{p_{\bot}}{E}}; \quad
    \Pi = \zeta \sqrt{m^{2} + p^{2}}.
\end{equation*}
Formula (\ref{eq:c_vector}) is valid for all $ n $ provided that one
assumes $ \zeta = \epsilon $ when $ n = 0 $. This has a physical
reason since the quantity $ \Pi $ is an eigenvalue of the spin
operator $ \g^{5} (P_{3} - \g^{3} m) $ which commutes with $ \HD $,
and $ \Pi = \epsilon \sqrt{m^{2} + p^{2}} $ at the lowest Landau
level. 

Now that we have found the energy spectrum and a system of
wavefunctions for Hamiltonian (\ref{eq:H_electron}), we may use the
perturbation theory to take into account possible small deviations of
$ {\bf b} $ from the direction of the magnetic field. Let $ {\bf b} =
(b_{\bot}, 0, b) $, the corresponding correction to $ \HD $ being $ V
= -\Sigma_{1} b_{\bot} $. There is no first-order correction to the
energy due to the rotational symmetry of the system. The second-order
correction obtained through the standard procedure is as follows: 
\newcommand{\rb}[1]{\raisebox{1.25ex}{#1}} % This is used to raise substacks in eq:delta_E_raw and eq:E_delta_E
\begin{equation}
\label{eq:delta_E_raw}
	\setlength\arraycolsep{0pt}
	\Delta E_{\bot} =
	\left. \left(
		\left. \frac{R_{+} R'_{-}}{E - E'} \right|_\text{\scriptsize\rb{$
			\begin{array}{lll}
				n' &=& n \! + \! 1 \\[-1pt]
				\epsilon' &=& \epsilon
			\end{array}
		$}}
		+
		\left. \frac{R_{+} R'_{-}}{E - E'} \right|_\text{\scriptsize\rb{$
			\begin{array}{lll}
				n' &=& n \! + \! 1 \\[-1pt]
				\epsilon' &=& -\epsilon
			\end{array}
		$}}
		+
		\left. \frac{R'_{+} R_{-}}{E - E'} \right|_\text{\scriptsize\rb{$
			\begin{array}{lll}
				n' &=& n \! - \! 1 \\[-1pt]
				\epsilon' &=& \epsilon
			\end{array}
		$}}
		+
		\left. \frac{R'_{+} R_{-}}{E - E'} \right|_\text{\scriptsize\rb{$
			\begin{array}{lll}
				n' &=& n \! - \! 1 \\[-1pt]
				\epsilon' &=& -\epsilon
			\end{array}
		$}}
	\right) \right|_\text{\scriptsize\rb{$
		\begin{array}{lll}
			p' &=& p \\[-1pt]
			\zeta' &=& -\zeta
		\end{array}
	$}},
\end{equation}
where $ E $ is given by Eq. (\ref{eq:E}),
\begin{equation*}
	R_{\pm} = \sqrt{2} b_{\bot} \left(
    	1 \pm \zeta \epsilon \sqrt{1 - \frac{p^{2}_{\bot}}{{E}^{2}}}
   	\right),
\end{equation*}
and we have used a stroke symbol to denote that a quantity is a
function of the quantum number set $ \{ n'p'\zeta'\epsilon' \} $
instead of $ \{ np\zeta\epsilon \} $, the latter being fixed for a
given one-particle state. The numbers $ n' $, $ \epsilon' $ are
expressed through $ n $, $ \epsilon $ differently for each term in
Eq. (\ref{eq:delta_E_raw}) while one has $ p' = p $ and $ \zeta' =
-\zeta $ in the whole expression. The last two terms with $ n' = n - 1
$ are absent in the case $ n = 0 $ due to $ R_{-} = 0 $ provided that
$ \zeta = \epsilon $. 

Despite the emerging energy level degeneracy with respect to $ \zeta
$, Eq. (\ref{eq:delta_E_raw}) is valid in the limit $ b \rarr 0
$. It is easy to notice though that the terms with $ \epsilon' =
\epsilon $ suffer from divergence due to a level crossing possible for
states with adjacent $ n $ and opposite $ \zeta $, thus making the
result obtained not applicable in the corresponding region of the
parameter space, namely, when $ 4 b^{2} \left( m^{2} + p^{2} \right)
\simeq (eH)^{2} $ [if $ \zeta = \sgn b $ the first term in
Eq. (\ref{eq:delta_E_raw}) is divergent and if $ \zeta = -\sgn b $
such is the third]. To workaround this, one has to modify the method
of calculating $ \Delta E_{\bot} $ in that region. Using the
perturbation theory formalism for two near-degenerate levels with
energies $ E $, $ E' $ (see, e.g., Ref.~\cite{bib:book_LL3} for
details), we find new energy values $ E_{\pm} $ with a gap induced by
the perturbation $ V $: 
\begin{equation}
\label{eq:E_delta_E}
	E_{\pm} =
	\frac{1}{2}
	\left(
		E + E' \pm \sqrt{(E - E')^{2} + 4 R R'}
	\right), \quad
	R R' =
    \begin{cases}
        R_{+} R'_{-}, & n' = n + 1, \\
        R'_{+} R_{-}, & n' = n - 1. \\
    \end{cases}
\end{equation}
Assuming $ E_{\pm} = E + \Delta E_{\bot} $ and taking into account that
\begin{equation*}
	\sgn(E - E') \sqrt{(E - E')^{2} + 4 R R'} \simeq
	E - E' + \frac{2RR'}{E - E'}
\end{equation*}
when $ RR' \ll |E - E'| $, Eqs. (\ref{eq:delta_E_raw}) and
(\ref{eq:E_delta_E}) can be combined into one asymptotic formula with
the change 
\begin{equation*}
	\frac{RR'}{E - E'} \rarr
	\frac{1}{2}
	\left(
		-E + E' + \sgn(E - E') \sqrt{(E - E')^{2} + 4 R R'}
	\right)
\end{equation*}
applied to the first and the third term in Eq. (\ref{eq:delta_E_raw}),
with $ R R' $ being $ R_{+} R'_{-} $ and $ R'_{+} R_{-} $
respectively. The factor $ \sgn(E - E') $ is used in the latter
expression to select a proper branch of solution (\ref{eq:E_delta_E})
since we want to retain the meaning of $ \Delta E_{\bot} $ being a
small correction to a particular energy level $ E $ when $ |E - E'| $
is not vanishing. There is an ambiguity in this approach arising when
$ E = E' $ in either of the terms which have undergone the change, it may be
fixed with the help of the following convention: 
\begin{equation*}
	\sgn(E - E') \rarr
    \begin{cases}
        \sgn_{+}(E - E'), & n' = n + 1, \\
        \sgn_{-}(E - E'), & n' = n - 1, \\
    \end{cases}
\end{equation*}
where $ \sgn_{\pm}(0) = \pm 1 $. It is easy to see that this ensures
the consistency of the formula obtained (no values of $ \Delta
E_{\bot} $ have been lost when considering the energy spectrum as a
whole). The final result reads
\begin{eqnarray}
	\Delta E_{\bot} =
	\Bigg [
	\setlength\arraycolsep{0pt}
	\left. \frac{1}{2} \left(
		-E + E' + \sgn_{+}(E - E') \sqrt{(E - E')^{2} + 4 R_{+} R'_{-}}
	\right) \right|_\text{\scriptsize\rb{$
		\begin{array}{lll}
			n' &=& n \! + \! 1 \\[-1pt]
			\epsilon' &=& \epsilon
		\end{array}
	$}}
	+
	\left. \frac{R_{+} R'_{-}}{E - E'} \right|_\text{\scriptsize\rb{$
		\begin{array}{lll}
			n' &=& n \! + \! 1 \\[-1pt]
			\epsilon' &=& -\epsilon
		\end{array}
	$}}
\nonumber \\
\label{eq:delta_E}
	+
	\setlength\arraycolsep{0pt}
	\left. \frac{1}{2} \left(
		-E + E' + \sgn_{-}(E - E') \sqrt{(E - E')^{2} + 4 R'_{+} R_{-}}
	\right) \right|_\text{\scriptsize\rb{$
		\begin{array}{lll}
			n' &=& n \! - \! 1 \\[-1pt]
			\epsilon' &=& \epsilon
		\end{array}
	$}}
	+
	\left. \frac{R'_{+} R_{-}}{E - E'} \right|_\text{\scriptsize\rb{$
		\begin{array}{lll}
			n' &=& n \! - \! 1 \\[-1pt]
			\epsilon' &=& -\epsilon
		\end{array}
	$}}
	\Bigg ] \Bigg |_\text{\scriptsize\rb{$
		\begin{array}{lll}
			p' &=& p \\[-1pt]
			\zeta' &=& -\zeta
		\end{array}
	$}}.
\end{eqnarray}
The spectrum $ E + \Delta E_{\bot} $ with $ E $ and $ \Delta E_{\bot}
$ provided with Eqs. (\ref{eq:E}) and (\ref{eq:delta_E}) can now be
used to evaluate the effective action of the model. 

% ======================================

\section{Effective Potential and Regularization\label{sec:pot}}

Let us now return to model (\ref{eq:L}). The corresponding one-loop effective action
\begin{equation}
\label{eq:Gamma}
	\Gamma =
	\int d^{4} x \left( -\frac{m^{2}}{4G} \right) +
	\frac{1}{\im} \ln \Det \left( \im \g D + \mu \g^{0} - m + \g^{5} \tau_{3} \g b \right)
\end{equation}
is decomposed trivially into similar parts calculated separately for
each flavor and color; moreover, it can be expressed in terms of the
effective action $ \Gamma' $ for the model studied in the previous
section with an appropriate change in the electric charge
and the chemical potential included:
\begin{eqnarray}
    \Gamma &=& 
    \int d^{4} x \left( -\frac{m^{2}}{4G} \right) + 
    N_{c} \, \Gamma' \big |_{e \rarr \frac{2}{3} e} +
    N_{c} \, \Gamma' \big |_{e \rarr \frac{1}{3} e},
\nonumber \\
\label{eq:Gamma_e}
	\Gamma' &=&
	\frac{1}{\im} \ln \Det \left( \im \g D + \mu \g^{0} - m - \g^{5} \g b \right) =
	\frac{1}{\im} \ln \Det \left( \im \pd^{0} + \mu - \HD \right),
\end{eqnarray}
where $ \HD $ is given in Eq. (\ref{eq:H_electron}). We have used a
charge conjugation for the up-quark when deriving the foregoing. Since
we know the eigenfunctions $ \{ \Psi_{nqp\zeta\epsilon} \} $ and the
spectrum $ \{ E_{np\zeta\epsilon} \} $ of $ \HD $, the expression for
$ \Gamma' $ can be evaluated through the standard procedure: 
\begin{eqnarray*}
	\Gamma'
	&=&
    \frac{1}{2 \im} \Tr \ln \left[ -(\im \pd^{0})^{2} + (\HD - \mu)^{2} \right]
\\
	&=&
	\frac{1}{2 \im} \int dp^{0} \sum_{(n)} \int d^{4}x \,
	\frac{1}{\sqrt{2 \pi}} \, e^{\im p^{0} t} \, \Psi^{+} \,
	\ln \left[ -(\im \pd^{0})^{2} + (\HD - \mu)^{2} \right]
	\frac{1}{\sqrt{2 \pi}} \, e^{-\im p^{0} t} \, \Psi
\\
	&=& 
	\frac{1}{2 \im} \int dp^{0} \sum_{(n)}
    \ln \left[ -(p^{0})^{2} + (E - \mu)^{2} \right]
    \frac{L_{t}}{2 \pi} \frac{L_{z}}{2 \pi} \frac{L_{y}}{2 \pi},
\end{eqnarray*}
where we have introduced a characteristic four-volume $ L_{t} L_{x} L_{y} L_{z} $ and
\begin{equation*}
    \sum_{(n)} \equiv \sum_{n \zeta \epsilon} \int dp \int dq =
    \sum_{n \zeta \epsilon} \int dp \, eH L_{x}.
\end{equation*}
In order to obtain the thermodynamic potential $ \Omega $, we employ
Matsubara's technique~\cite{bib:Matsubara}: 
\begin{equation*}
	\int\limits_{-\infty}^{\infty} \frac{dp^{0}}{2 \pi} \rarr
	\int\limits_{-\im \infty}^{\im \infty} \frac{dp^{0}}{2 \pi} =
	\im \int\limits_{-\infty}^{\infty} \frac{dp^{4}}{2 \pi} \rarr
	\im \frac{1}{\beta} \sum\limits_{k=-\infty}^{+\infty}, \quad
	p^{0} \rarr \im p^{4} \rarr \im \omega_{k} = \im \frac{2 \pi}{\beta} \left( k + \frac{1}{2} \right),
\end{equation*}
where $ \beta = 1 / T $ is the inverse temperature; the sum over $ k $
is easily evaluated. We finally find
\begin{eqnarray} 
    \Omega &=&
    -\frac{\Gamma}{L_{t} L_{x} L_{y} L_{z}} =
    \frac{m^{2}}{4G} +
    N_{c} \, \Omega' \big |_{e \rarr \frac{2}{3} e} +
    N_{c} \, \Omega' \big |_{e \rarr \frac{1}{3} e},
\nonumber \\
\label{eq:Omega_e}
    \Omega' &=&
    -\frac{1}{2} \frac{eH}{(2 \pi)^{2}}
    \int dp \sum_{n \zeta \epsilon}
    \left[
        | E - \mu | + \frac{2}{\beta} \ln \left( 1 + e^{-\beta | E - \mu |} \right)
    \right].
\end{eqnarray}

Separating the effects of nonzero temperature and the vacuum
contribution, expression (\ref{eq:Omega_e}) can be decomposed into
three terms: 
\begin{equation*}
	\Omega' = \Omega'_{\rm v} + \Omega'_{\mu} + \Omega'_{T},
\end{equation*}
where
\begin{eqnarray}
\label{eq:Omega_e_v}
    \Omega'_{\rm v} &=&
    -\frac{1}{2} \frac{eH}{(2 \pi)^{2}}
    \int dp \sum_{n \zeta \epsilon}
    | E |,
\\
\label{eq:Omega_e_mu}
    \Omega'_{\mu} &=&
    -\frac{1}{2} \frac{eH}{(2 \pi)^{2}}
    \int dp \sum_{n \zeta \epsilon}
    \left( | E - \mu | - | E | \right),
\\
\label{eq:Omega_e_T}
    \Omega'_{T} &=&
    -\frac{1}{\beta} \frac{eH}{(2 \pi)^{2}}
    \int dp \sum_{n \zeta \epsilon}
    \ln \left( 1 + e^{-\beta | E - \mu |} \right).
\end{eqnarray}
The vacuum term $ \Omega'_{\rm v} $ is divergent while the terms $
\Omega'_{\mu} $ and $ \Omega'_{T} $ are finite (being zero when $ \mu
= 0 $ and $ T = 0 $ respectively). The NJL model is known to be
sensitive to the choice of a regularization scheme due to the
nonrenormalizable nature of the four-fermion interaction
\cite{bib:NJL_reg_dep} (see also a discussion on this subject for
the case of a spatially nonuniform condensate in
Ref.~\cite{bib:Broniowski_Kutschera_amb}). We here employ the
proper-time method~\cite{bib:ptr}: 
\begin{equation*}
    \Omega'_{\rm v} \rarr
    \frac{1}{4 \sqrt{\pi}} \frac{eH}{(2 \pi)^{2}}
    \int dp \sum_{n \zeta \epsilon}
    \int\limits_{1/\Lambda^{2}}^{+\infty} \frac{ds}{s^{3/2}} \, e^{-s E^{2}},
\end{equation*}
where $ \Lambda $ is the regularization parameter; so our results
should agree with those obtained in
Ref.~\cite{bib:Nakano_Tatsumi_dcdw} in the limit~$ H \rarr 0 $. 

It is easy to see that expression (\ref{eq:Omega_e_T}) for $
\Omega'_{T} $ is well defined, and although expression
(\ref{eq:Omega_e_mu}) for $ \Omega'_{\mu} $ seems to be convergent
due to the internal sum over $ \epsilon $,
care is needed when evaluating it since it has been obtained as a
difference of two divergent objects. One can derive an arbitrary value
for such an expression rearranging the terms during the summation
procedure~\cite{bib:div_sum}, so an intermediate regularization is
needed to get a correct result. Let it be a simple cutoff: 
\begin{equation}
\label{eq:Omega_e_mu_reg}
    \Omega'_{\mu} \rarr
    -\frac{1}{2} \frac{eH}{(2 \pi)^{2}}
    \int dp \sum_{n \zeta \epsilon}
    \left( | E - \mu | - | E | \right) \, \theta(\Lambda' - |E|),
\end{equation}
where $ \Lambda' $ is sufficiently large (not necessarily being equal
to $ \Lambda $). If the relation $ E|_{\epsilon = +1} = -E|_{\epsilon
= -1} > 0 $ holds, the cutoff factor can be dropped out (provided
that $ \mu < \Lambda' $): 
\begin{eqnarray*}
	&&
	\sum_{\epsilon} \left( | E - \mu | - | E | \right) \, \theta(\Lambda' - |E|)
	= \left.  \big [ \left( | E - \mu | + E + \mu - 2 E \right) \, \theta(\Lambda' - E) \big ] \right|_{\epsilon = +1}
\\
    &&
    = \left. \big [ 2 \left( \mu - E \right) \, \theta(\mu - E) \, \theta(\Lambda' - E) \big ] \right|_{\epsilon = +1}
    = \left. \big [ 2 \left( \mu - E \right) \, \theta(\mu - E) \big ] \right|_{\epsilon = +1}.
\end{eqnarray*}
But this is not the case when the symmetry between the particle and
antiparticle spectra is broken: $ E|_{\epsilon = +1} \neq
-E|_{\epsilon = -1} $, which occurs at the lowest Landau level in our
problem [and when taking into account corrections (\ref{eq:delta_E})
to the energy levels as well]. In general, one has to retain the
regularization throughout the calculations or modify the whole
expression by a finite but nonzero correction. The effect for the
case $ b_{\bot} = 0 $, $ n = 0 $ can be studied exactly (see the Appendix): 
\begin{equation*}
    \int dp \sum_{\epsilon}
    \left( | E - \mu | - | E | \right) \, \theta(\Lambda' - |E|) \bigg|_{\Lambda' \rarr \infty} = 
    \int dp \sum_{\epsilon}
    \left( | E - \mu | - | E | \right) + 4 \mu b.
\end{equation*}
If one omits the term $ 4 \mu b $ in the above expression, the
resulting potential $ \Omega $ turns to be dependent on $ b $ when $ m
= 0 $, and this is physically incorrect according to definition
(\ref{eq:cond_def}); no observable quantity may depend on the
wave vector of a condensate wave with a zero amplitude. 

% ======================================

\section{Phase Diagram\label{sec:diag}}

To construct a phase diagram of the system, we have studied the minima
of the regularized thermodynamic potential $ \Omega $ numerically with
respect to the order parameters $ m $ and $ b $ for different values
of the chemical potential $ \mu $ and the magnetic field strength $ H
$. We tried to find a global minimum in the case of several minima present
on the $ \Omega $ surface. We used spectrum (\ref{eq:E}) and took into
account corrections (\ref{eq:delta_E}) to study the stability of the
results with respect to small deviations of $ {\bf b} $ from the
direction of the magnetic field (taking $ \Lambda' = 10 \, \Lambda
$). We only used dimensionless quantities throughout the calculations
with $ \Lambda $ being the characteristic energy scale. In what follows, we denote
these quantities with the same symbols as the original ones, e.g., $ m
$ stands for $ m / \Lambda $, etc. We performed integration over
the quantum number $ n $ instead of summation when $ eH \ll 1 $, thus
being able to consider the limit $ H \rarr 0 $ with no
singularities. The estimate of the maximum relative and absolute error
was set at the level of $ 10^{-3} $ and $ 10^{-8} $, respectively.
We take the values of $ \mu $
and $ \sqrt{eH} $ from $ 0 $ up to $ 0.8 $; it should be noted that
there is no physical sense in considering high values of these
parameters since $ 1 $ is the (dimensionless) regularization constant
in our model.
The critical value of the coupling constant is $ G_{c} \simeq 3.27 $ in our model.
If $ H = 0 $ and $ \mu = 0 $,
spontaneous chiral symmetry breaking only occurs when $ G > G_{c} $;
we take this fact as the definition of $ G_{c} $.

At present, an exact form of the one-particle energy spectrum in the case of $ b_{\bot} > 0 $ ($ m > 0 $) is not known,
so that comprehensive analysis of the problem cannot be made.
There is no strict guarantee that there are no global minima of the thermodynamic potential somewhere in the region
$ m > 0 $, $ b > 0 $, $ b_{\bot} > 0 $ when $ H > 0 $, $ \mu > 0 $
since one can construct a dimensionless ratio $ \sqrt{eH} / \mu $
and the latter may be related to the ratio $ b_{\bot} / b $.
If this is true, the DCDW wave vector orientation would be diverted from the preferred direction of $ {\bf H} $
and the rotational symmetry of the system would be completely broken.
Nonetheless, it is reasonable to believe that the global minima of $ \Omega $ are reached when $ b_{\bot} = 0 $
implying that the rotational symmetry is still preserved.
To test this to the extent possible, for each minimum found (when $ H > 0 $),
we studied the behavior of the thermodynamic potential in the region of $ b_{\bot} $ close to zero.
We actually calculated the second derivative
$ \partial^{2} \Omega / \partial b_{\bot}^{2} |_{b_{\bot} \rarr 0} $ numerically,
and we made use of the explicit energy spectrum corrections~(\ref{eq:delta_E})
during the evaluation of that quantity.
The latter turned to be positive everywhere,
so, in this approximation,
no instability of the thermodynamic potential minima with respect to $ b_{\bot} $ has been found.

\begin{figure*}
\includegraphics{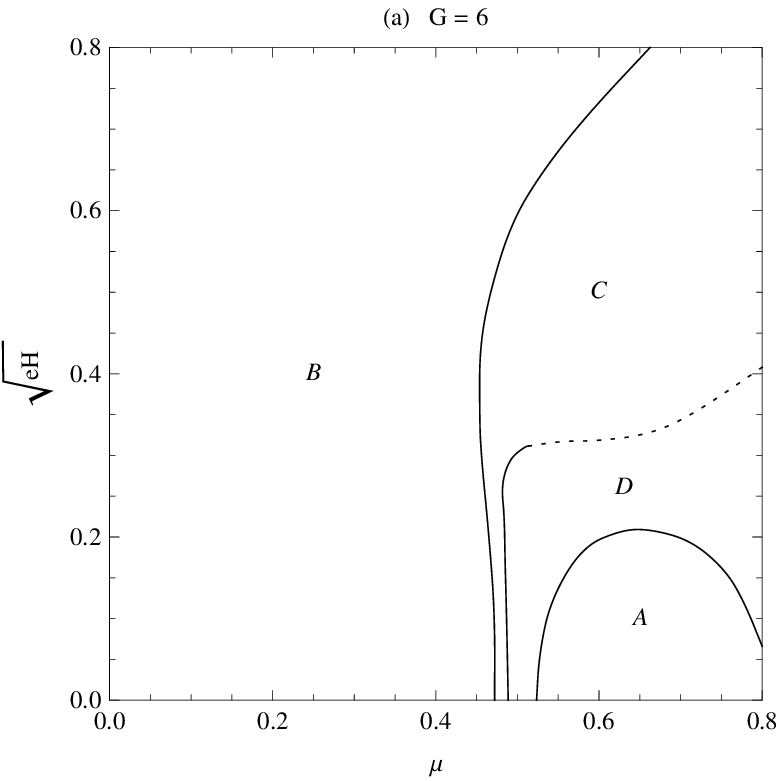}
\quad 
\includegraphics{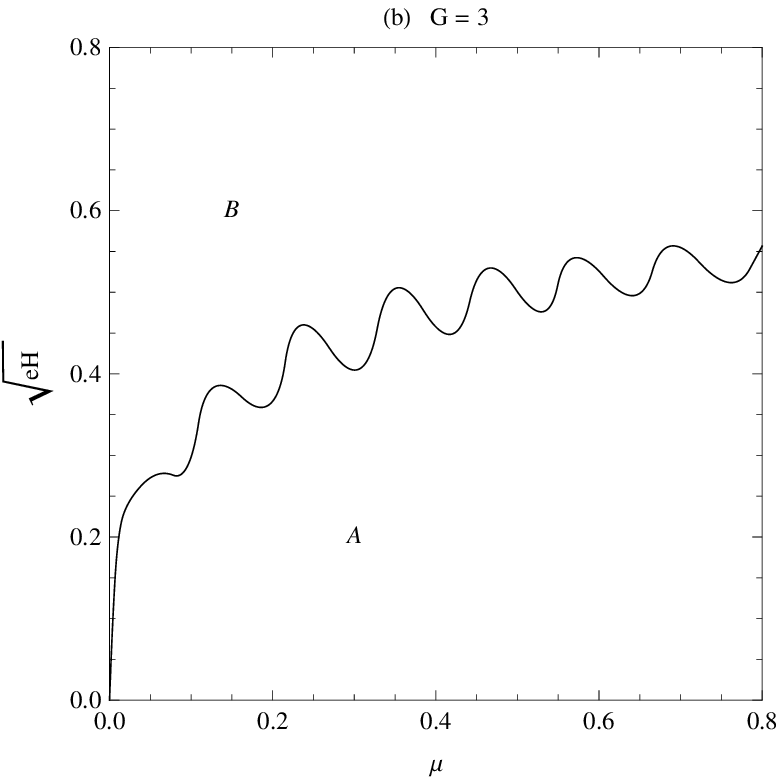}
\caption{Phase diagrams for (a) supercritical and (b) subcritical values of
the coupling constant $ G $ at zero temperature. All quantities are
dimensionless.
There is a symmetric massless phase $ A $ with no chiral condensate
and two chirally broken massive phases $ B $ and $ C $,
the latter being a phase with a nonzero matter density $ \rho $,
whereas $ \rho = 0 $ in phase $ B $.
Massive phases $ B $ and $ C $ are spatially nonuniform when $ H > 0 $.
There is also a new phase $ D $ with a strong condensate inhomogeneity
(retaining the presence of DCDW in the $ H \rarr 0 $ limit).
The phase transitions are first order.
There is a crossover between phases $ C $ and $ D $ in a magnetic field strong enough;
we have plotted the boundary between them
with a dotted line in that region.
\label{fig:phases}}
\end{figure*}

\begin{figure*}
\includegraphics{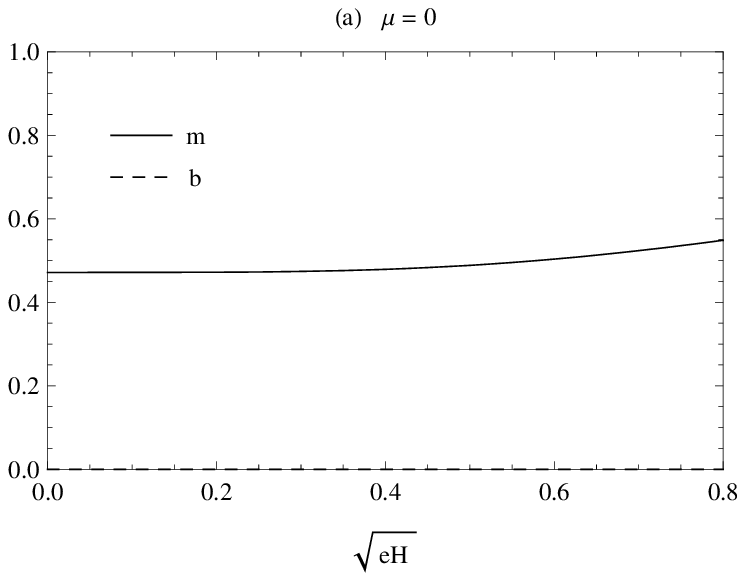}
\quad 
\includegraphics{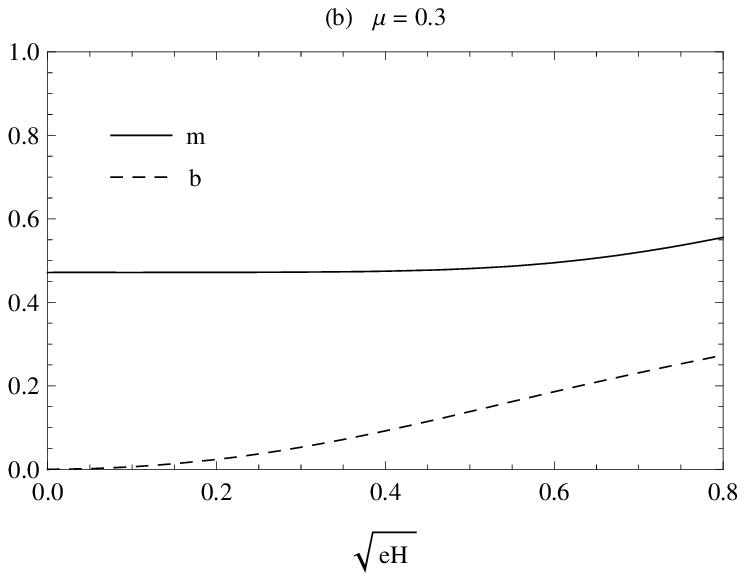}
\vskip 15pt
\includegraphics{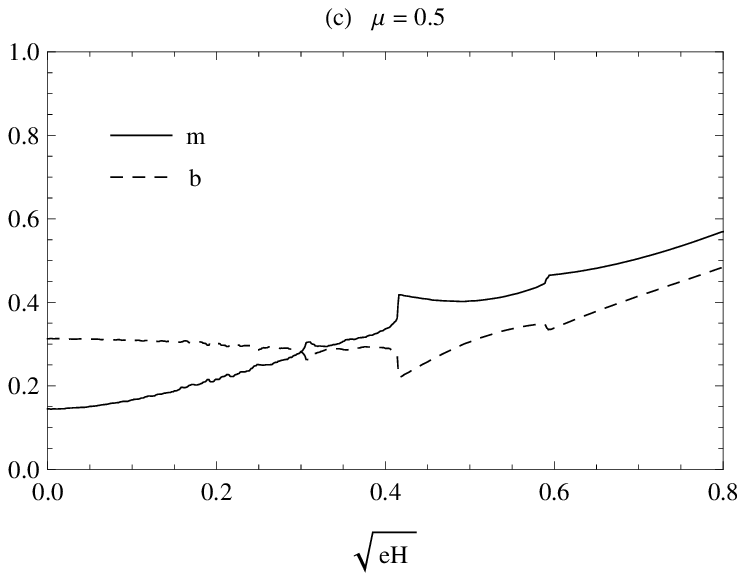}
\quad 
\includegraphics{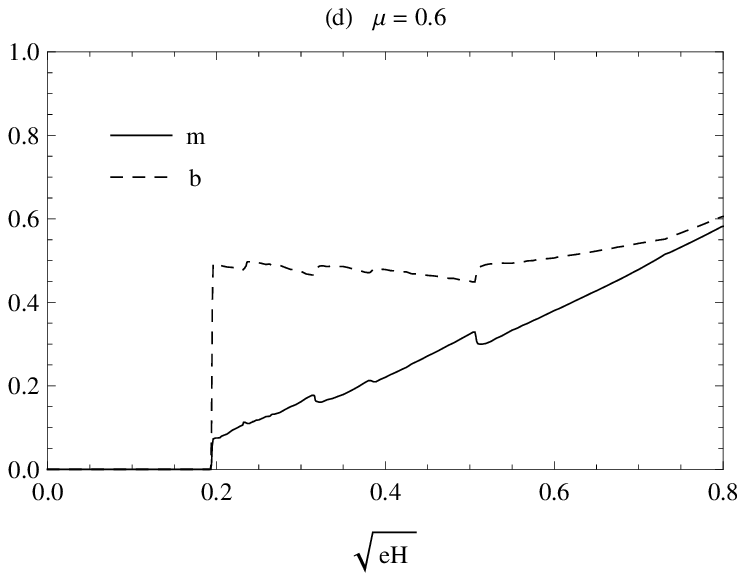}
\caption{The order parameters as functions of the magnetic field strength
for various values of the chemical potential at zero temperature, 
and $ G = 6 $. All quantities are dimensionless.\label{fig:mb6x}}
\end{figure*}

\begin{figure*}
\includegraphics{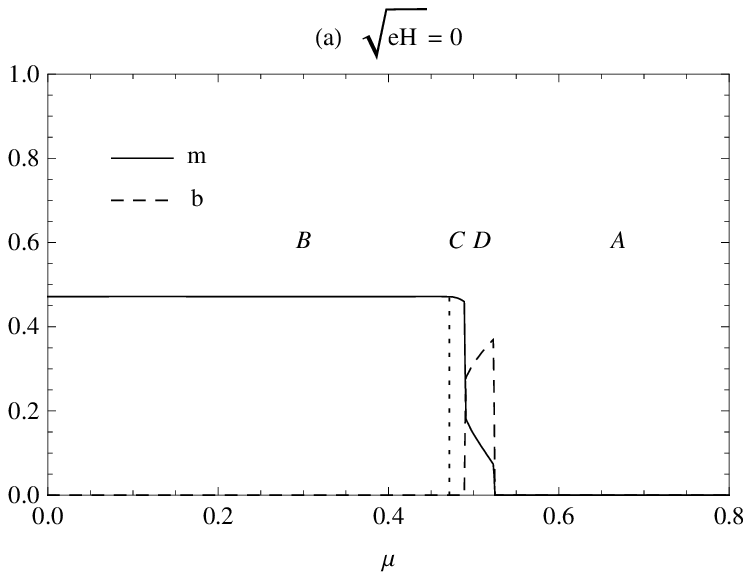}
\quad 
\includegraphics{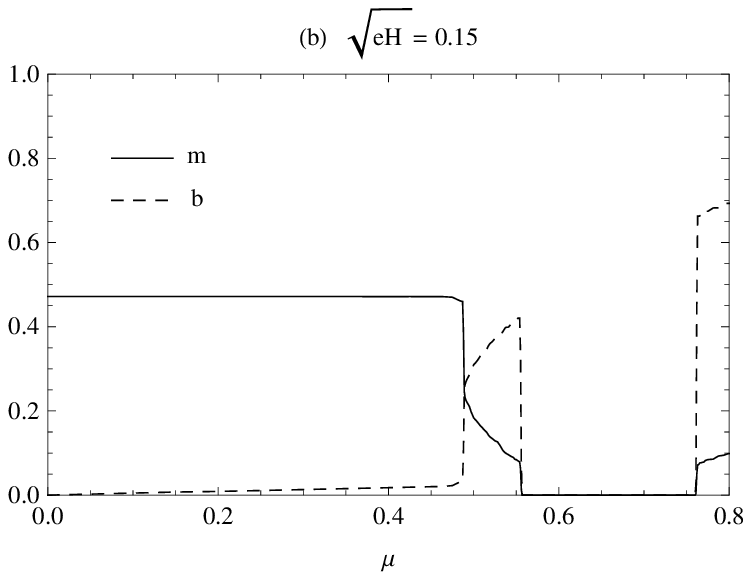}
\vskip 15pt
\includegraphics{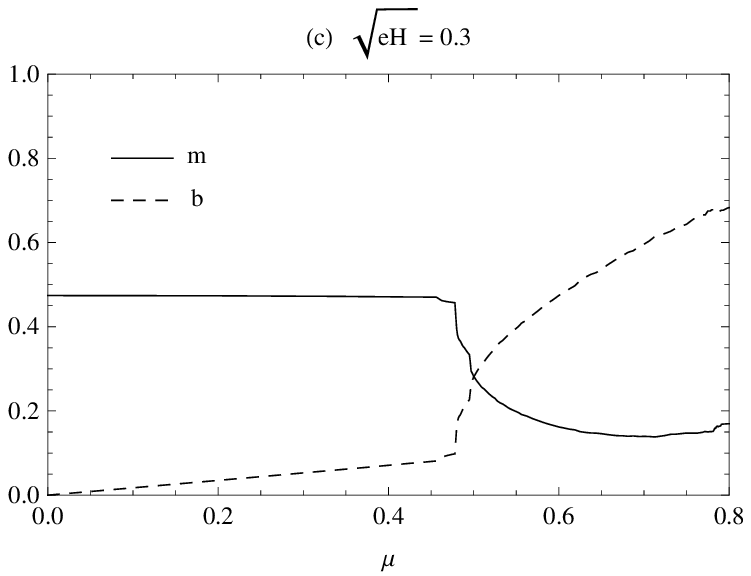}
\quad 
\includegraphics{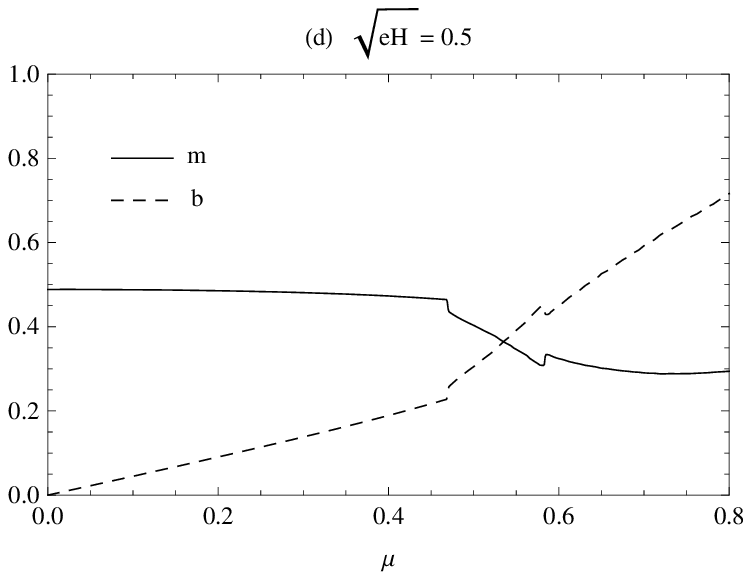}
\caption{The order parameters as functions of the chemical potential
for various values of the magnetic field strength at zero temperature, 
and $ G = 6 $. All quantities are dimensionless.\label{fig:mb6y}}
\end{figure*}

\begin{figure*}
\includegraphics{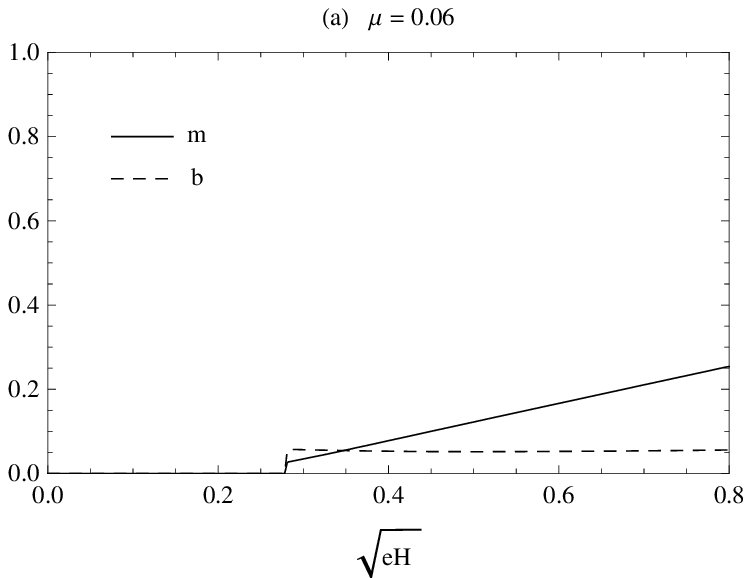}
\quad 
\includegraphics{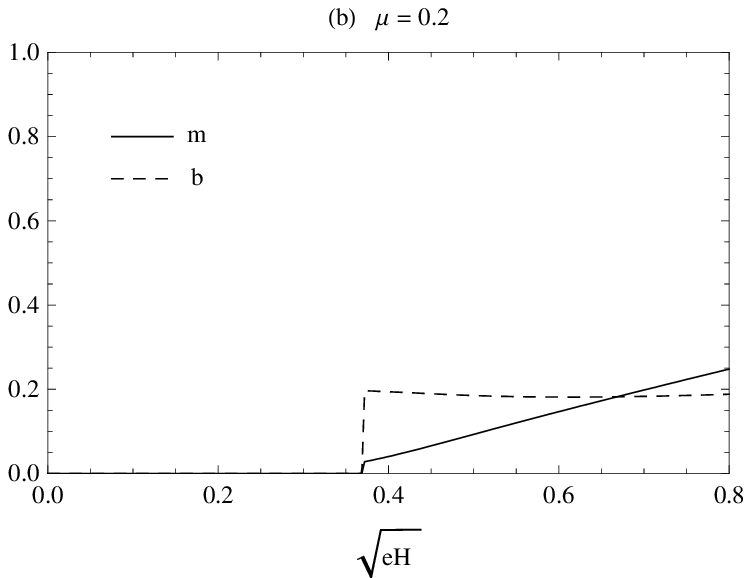}
\vskip 15pt
\includegraphics{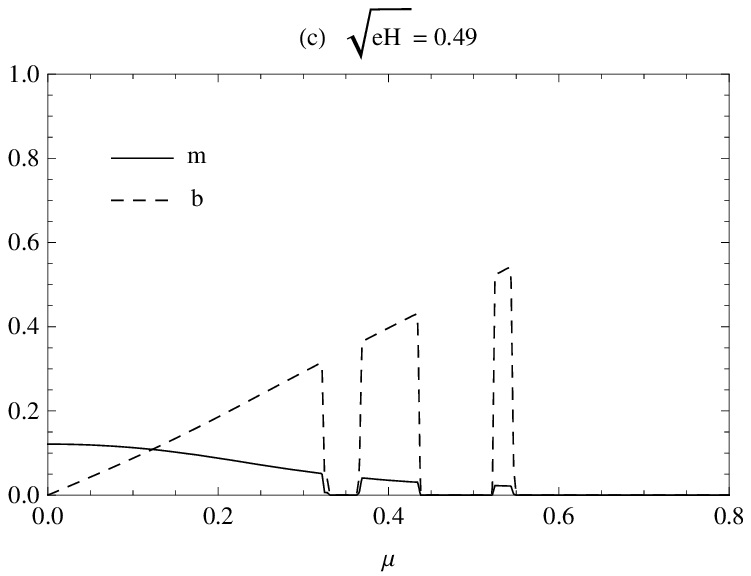}
\quad 
\includegraphics{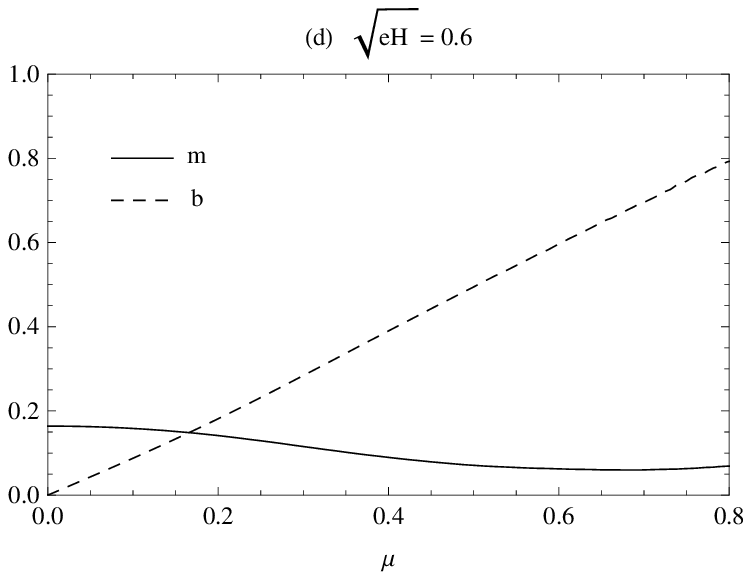}
\caption{Samples of the order parameter dependence on the external
conditions at zero temperature, $ G = 3 $. All quantities are
dimensionless.\label{fig:mb3}} 
\end{figure*}

The results of numerical analysis in the case of $ T = 0 $ and supercritical $ G = 6 $
are presented in Figs. \ref{fig:phases}a, \ref{fig:mb6x}, \ref{fig:mb6y}.
As one would expect, we recover the result obtained in
Ref.~\cite{bib:Nakano_Tatsumi_dcdw} in the limit $ H \rarr 0 $, see
Fig. \ref{fig:mb6y}(a); and there is a nontrivial behavior of the
system when $ H > 0 $. The order parameter $ b $ related to the DCDW
wave vector ($ b = q / 2 $) grows either smoothly (for the range of the
chemical potential $ \mu $ up to some value) or discontinuously (for
higher values of $ \mu $) with the increase of the magnetic field
strength $ H $, the effect being more vivid for greater $ \mu $, see
Fig. \ref{fig:mb6x}. There is also a gap corresponding to a transition
from a symmetric phase present in a weak field in dense matter (we
assume $ b = 0 $ when $ m = 0 $ although $ b $ has no physical meaning
in that case and can be set to have an arbitrary value). DCDW is absent if
$ \mu = 0 $. Noticeably, the order parameter $ b $ grows linearly with
the increase of the chemical potential $ \mu $ (up to a critical value
where a phase transition occurs), the growth rate being higher in the
stronger field, see Fig. \ref{fig:mb6y}. DCDW is absent when $ H = 0 $
except for the range of $ \mu $ corresponding to a new phase examined
in Ref.~\cite{bib:Nakano_Tatsumi_dcdw}. This phase (we name it phase $
D $, see below) undergoes further development with the increase of the
magnetic field strength forcing out the symmetric phase with $ m = 0
$. The order parameter oscillation visible in the diagrams when the
chemical potential is high enough is a phenomenon typical for the
model~\cite{bib:VVK, bib:VKE, bib:EKVV}, such behavior is generally
inherent in cold many-body quantum systems in a magnetic field, the
fact known since the studies on the de~Haas--van~Alphen
effect~\cite{bib:de_Haas_van_Alphen, bib:Landau_dia}. 

In general, the NJL model is known to give rise to three
distinct phases~\cite{bib:NJL_ABC_1, bib:NJL_ABC_2}:
a symmetric massless phase $ A $ with no chiral condensate
and two chirally broken massive phases $ B $ and $ C $,
the latter being a phase with a nonzero matter density $ \rho $,
whereas $ \rho = 0 $ in phase $ B $
(the $ B \rarr C $ transition occurs for $ \mu > m $).
Massive phases $ B $ and $ C $ are now spatially nonuniform when $ H > 0 $.
There is also a new phase with a strong condensate inhomogeneity
(retaining the presence of DCDW in the $ H \rarr 0 $ limit studied in
Ref.~\cite{bib:Nakano_Tatsumi_dcdw}), we denote it with the symbol $ D $.
The position of the phases described above in case of $ H = 0 $ is
illustrated in Fig. \ref{fig:mb6y}(a)
and their evolution with the increase of the magnetic field strength is shown in Fig. \ref{fig:phases}(a).
The transitions between the phases under consideration are first order
since the order parameters are discontinuous
except for the $ B \rarr C $ transition when $ H = 0 $
which is second order being a singular point in the diagram \cite{bib:EKVV}.
It should be noted that the $ D \rarr C $ transition occurring in a magnetic field strong enough
actually belongs to a series of typical order parameter oscillations visible, e.g., in Figs. \ref{fig:mb6x}c, \ref{fig:mb6x}d.
There is no significant physical difference between $ C $ and $ D $ in that region
so we consider it as a crossover area plotting the corresponding transition with a dotted line,
and we only plot a solid line between $ C $ and $ D $ in the region
where these phases can be distinguished clearly with a noticeable change in their physical properties
[see, e.g., Figs. \ref{fig:mb6y}a, \ref{fig:mb6y}b].
The position of the end point separating the solid and the dotted segment
is therefore not fixed precisely and should be chosen judiciously.
In the most general case,
one may consider an infinite
series of phases $ \{ A_{n} \} $, $ \{ C_{n} \} $, and $ \{ D_{n} \} $
when a magnetic field is present with phase transitions corresponding
to the order parameter oscillations mentioned above (like it is
done in Refs.~\cite{bib:VVK, bib:VKE, bib:EKVV},
see also a recent study in Ref. \cite{bib:HCMF6}).
However, since these oscillations
are small in their relative magnitude and
tend to be smeared out with finite temperature taken into account, 
we consider such series as single phases,
and in this reasonable approximation, this situation may be treated as
a crossover between $ C $ and $ D $. 
The main result we have obtained is that there is a
nonzero $ b $ in all phases when $ H > 0 $ except for the symmetric
one and the case of $ \mu = 0 $. Smooth and linear growth of $ b $
with the increase of $ H $ and $ \mu $ is inherent in phase $ B
$. Symmetric phase $ A $ now occupies a limited area on the
diagram.

We have also examined the case of a subcritical $ G $ in addition to
the strong-coupling regime. The results for $ G = 3 $, $ T = 0 $ are
presented in Figs. \ref{fig:phases}b, \ref{fig:mb3}. The magnetic field
is known to be a catalyst of the spontaneous chiral symmetry breaking
both in renormalizable and nonrenormalizable (NJL-like) theories
(see, e.g., Refs.~\cite{
bib:Klimenko_GN_1, bib:Klimenko_GN_2, bib:Klimenko_GN_3, bib:VMZK,
bib:GMS1, bib:GMS2, bib:GMS3, bib:Gorbar_HD,
bib:SSW, bib:Incera} and also Refs.~\cite{bib:VVK, bib:VKE, bib:EKVV}), the
latter demonstrating the emergence of a dynamic fermion mass for
arbitrary small values of the coupling constant. This effect is
present in our case as well. The phase diagram structure obtained for
our model is similar to that derived in Ref.~\cite{bib:EKVV} for $ G <
G_{c} $ and phase $ B $ exhibits the same behavior of the order
parameter~$ b $ growth as described above. Thus, DCDW formation is
preferable for the system in a wide range of the coupling constant. 

% ======================================

\section{Conclusions\label{sec:outro}}

The calculations performed in the framework of the NJL model have
shown that the presence of an external magnetic field favors the
formation of a spatially nonuniform chiral condensate (in the form of
a dual chiral density wave) in dense quark matter at low
temperatures.
This means that there exists a critical magnetic field strength $ H_{c} $
such that one of the spatially nonuniform DCDW phases ($ B $, $ C $, or $ D $)
emerges in the system when $ H > H_{c} $
both for supercritical and subcritical values of the coupling constant
for arbitrary nonzero values of the chemical potential $ \mu $ and $ T = 0 $.
For example, if $ G > G_{c} $ then it is easy to see that
$ H_{c} > 0 $ for the range of $ \mu $ corresponding to the symmetric phase~$ A $,
whereas $ H_{c} = 0 $ for other values of $ \mu $ [see Fig. \ref{fig:phases}(a)].
On the contrary, in the case of subcritical $ G < G_{c} $,
the quantity $ H_{c} $ is nonvanishing for all $ \mu > 0 $ [see Fig. \ref{fig:phases}(b)].
One can verify that the effect of the chiral condensate spatial modulation
is mainly due to the
particle and antiparticle energy spectra asymmetry induced by the
presence of DCDW in our model; if one drops out the contribution of
the distorted lowest Landau level (LLL)
to the thermodynamic potential of the
system, the phenomenon of the condensate wave vector being nonzero in
the massive phases of the model except for $ D $ will be lost and
phase $ D $ will be far less stable occupying a small area on the
phase diagram.

As discussed in Ref.~\cite{bib:Nakano_Tatsumi_dcdw} (see also
Ref.~\cite{bib:Ohwa}), linear growth of the condensate wave
vector with the increase of the chemical potential is generally inherent in
one-dimensional systems and this is in agreement with the dimensional
reduction phenomenon occurring for fermions in a strong magnetic field
\cite{bib:GMS2, bib:GMS3, bib:Gorbar_HD} (see also Ref.~\cite{bib:EbZh_dim_red}
for the case of chromomagnetic fields);
this behavior of the order parameters is actually related to the specific properties of the~LLL.
A singular role of the~LLL and its impact on physical phenomena in various problems
concerning dense matter and symmetry breaking
is pointed out in a number of studies,
see, e.g., Ref. \cite{bib:Miransky_spectrum}
and also a recent discussion on the chiral magnetic effect in Ref. \cite{bib:Basar_spirals}.

In this paper, we have only 
reported our results for the case of cold quark matter,
and the role of finite temperature is to be studied in our forthcoming publications. 
There are also other interesting subjects left beyond
the scope of our paper. 
Since quark matter may possess its own
magnetization (see, e.g., Refs.~\cite{bib:quark_ferro_1,
bib:quark_ferro_2, bib:quark_ferro_3, bib:quark_ferro_4} and also
Refs.~\cite{bib:Son_Stephanov, bib:Noronha_Shovkovy}), a challenging
self-consistent problem may arise with the magnetic field being
generated dynamically. One should also consider the color
superconductivity phenomenon possible along with the chiral density
waves formation; the results obtained in such generalized models seem
to be less regularization dependent~\cite{bib:Partyka_Sadzikowski}. A
nonzero quark current mass should be taken into account as well. 
Besides, concerning the ground state spatial configuration of the NJL model,
it has been argued that
domain walls may be more preferable than chiral density waves
at least in the absence of external gauge fields \cite{bib:Nickel}.
At the same time, as it may be concluded from our calculations,
one would expect that a strong magnetic field favors the formation of DCDW. 
Thus, there should exist some solution interpolating between the two extremes
in the intermediate region of the magnetic field strength,
possibly being similar to the solution discussed in Ref. \cite{bib:Basar_combi}.
On the other hand,
a competing mechanism for domain walls formation in a strong magnetic field
has also been discussed in literature \cite{bib:Son_Stephanov}.
Therefore, the problem of the preferred ground state spatial structure
requires further theoretical investigation
but, in general, it has been shown
that an external magnetic field induces condensate inhomogeneity,
in the form of DCDW or some more preferable configuration.
Another subject of research is obtaining 
analytical expressions for the order parameters
as functions of
the external conditions
in a weak magnetic field at least in some special cases
using an approach similar to that adopted, e.g., in Ref. \cite{bib:Shushpanov}.

Our concluding remark is that real existence of a spatially
nonuniform chiral condensate in nature is yet an open question
since theoretical results related to the problem are generally model
and approximation dependent.
Unfortunately, exact QCD calculation of its production is impossible,
since this is an infrared phenomenon.  
Nonetheless, we believe that the theoretical research 
of this kind of nonperturbative effects
will yield our better understanding the properties of
strongly interacting matter. 

% ======================================

\appendix*
\section{Regularization of $ \Omega'_{\mu} $ at the $ n = 0 $ Level}

Let us consider the contribution of the $ n = 0 $ energy level in expression (\ref{eq:Omega_e_mu_reg}):
\begin{equation}
\label{eq:K_en}
	K^{\rm en} =
    \int\limits_{0}^{+\infty} dp \sum_{\epsilon}
    \left( | E - \mu | - | E | \right) \, \theta(\Lambda' - |E|), \quad
    E = \epsilon \sqrt{m^{2} + p^{2}} + b,
\end{equation}
where we have utilized the parity of $ E $ with respect to $ p $ under the integral and omitted the constant factor $ -eH / (2 \pi)^{2} $. One can easily prove the following formulas being valid for sufficiently large values of $ P $:
\begin{eqnarray*}
	\int\limits_{0}^{P} dp \, \left| \sqrt{m^{2} + p^{2}} + a \right| &=&
	\begin{cases}
		I(0,P) + Pa, & a > -|m|, \\
		-I(0,p_{0}) + I(p_{0},P) - 2p_{0}a + Pa, & a < -|m|,
	\end{cases} \\
	\int\limits_{0}^{P} dp \, \left| -\sqrt{m^{2} + p^{2}} + a \right| &=&
	\begin{cases}
		I(0,P) - Pa, & a < |m|, \\
		-I(0,p_{0}) + I(p_{0},P) + 2p_{0}a - Pa, & a > |m|,
	\end{cases}
\end{eqnarray*}
where $ p_{0} = \sqrt{a^{2} - m^{2}} $ and
\begin{equation*}
	I(p_{1}, p_{2}) \equiv
	\int\limits_{p_{1}}^{p_{2}} dp \, \sqrt{m^{2} + p^{2}} =
	\frac{p_{2}}{2} \sqrt{m^{2} + p_{2}^{2}} - \frac{p_{1}}{2} \sqrt{m^{2} + p_{1}^{2}} +
	\frac{m^{2}}{2} \ln \left( \frac{p_{2} + \sqrt{m^{2} + p_{2}^{2}}}{p_{1} + \sqrt{m^{2} + p_{1}^{2}}} \right).
\end{equation*}
Let us apply the above formulas to evaluate $ K^{\rm en} $. We have to choose different integration limits for different values of $ \epsilon $ due to the energy cutoff and the spectrum asymmetry: $ P_{1} = \sqrt{(\Lambda' - b)^{2} - m^{2}} $ for $ \epsilon = +1 $ and $ P_{2} = \sqrt{(\Lambda' + b)^{2} - m^{2}} $ for $ \epsilon = -1 $. Consider the following expression:
\begin{eqnarray}
	J^{\rm en} &=&
    \int\limits_{0}^{P_{1}} dp \,
    \left| \sqrt{m^{2} + p^{2}} + a \right| +
    \int\limits_{0}^{P_{2}} dp \,
    \left| -\sqrt{m^{2} + p^{2}} + a \right| 
\nonumber \\
    &=&
    \begin{cases}
    	I(p_{0},P_{1}) + I(p_{0},P_{2}) + (P_{1} - P_{2})a + 2p_{0}a, & |m|<a, \\
    	I(0, P_{1}) + I(0, P_{2}) + (P_{1} - P_{2})a, & -|m| < a < |m|, \\
    	I(p_{0},P_{1}) + I(p_{0},P_{2}) + (P_{1} - P_{2})a - 2p_{0}a, & a<-|m|. \\
    \end{cases}
\end{eqnarray}
Let us compare this result to one obtained with the help of a trivial momentum cutoff (i.e., without the factor $ \theta(\Lambda' - |E|) $ but with a common upper limit in the integrals). Assuming $ P_{1} = P_{2} = \ti\Lambda' $, we get
\begin{eqnarray}
	J^{\rm mom} &=&
    \int\limits_{0}^{\ti\Lambda'} dp \,
    \left| \sqrt{m^{2} + p^{2}} + a \right| +
    \int\limits_{0}^{\ti\Lambda'} dp \,
    \left| -\sqrt{m^{2} + p^{2}} + a \right|
\nonumber \\
	&=&
    \begin{cases}
    	2 I(p_{0}, \ti\Lambda') + 2p_{0}a, & |m|<a, \\
    	2 I(0, \ti\Lambda'), & -|m| < a < |m|, \\
    	2 I(p_{0}, \ti\Lambda') - 2p_{0}a, & a<-|m|. \\
    \end{cases}
\end{eqnarray}
The difference of the above quantities is
\begin{equation*}
	J^{\rm en} - J^{\rm mom} = I(\ti\Lambda',P_{1}) + I(\ti\Lambda',P_{2}) + (P_{1} - P_{2})a.
\end{equation*}
Thus we have
\begin{eqnarray*}
	K^{\rm en} &=&
	J^{\rm en}|_{a=b-\mu} - J^{\rm en}|_{a=b} = J^{\rm mom}|_{a=b-\mu} - J^{\rm mom}|_{a=b} + \mu(P_{2} - P_{1})
	= K^{\rm mom} + \mu(P_{2} - P_{1}),
\end{eqnarray*}
where we have introduced the symbol $ K^{\rm mom} $ to denote expression (\ref{eq:K_en}) regularized with a trivial momentum cutoff instead of the $ \theta(\Lambda' - |E|) $ factor:
\begin{equation}
\label{eq:K_mom}
	K^{\rm mom} =
    \int\limits_{0}^{\ti\Lambda'} dp \sum_{\epsilon}
    \left( | E - \mu | - | E | \right).
\end{equation}
Since
\begin{equation*}
	\lim_{\Lambda' \rarr \infty} \left( P_{2} - P_{1} \right) = 2b,
\end{equation*}
we finally get
\begin{equation}
    \int\limits_{0}^{+\infty} dp \sum_{\epsilon}
    \left( | E - \mu | - | E | \right) \, \theta(\Lambda' - |E|) \Bigg|_{\Lambda' \rarr \infty} = 
    \int\limits_{0}^{\ti\Lambda'} dp \sum_{\epsilon}
    \left( | E - \mu | - | E | \right) \Bigg|_{\ti\Lambda' \rarr \infty} + 2 \mu b.
\end{equation}
We should note that expression (\ref{eq:K_mom}) for $ K^{\rm mom} $ is convergent and well-defined in the limit $ \ti\Lambda' \rarr \infty $. Taking sufficiently large values of $ p $, one has
\begin{equation*}
    \sum_{\epsilon}
    \left(
    	\left| \epsilon \sqrt{m^{2} + p^{2}} + b - \mu \right| -
    	\left| \epsilon \sqrt{m^{2} + p^{2}} + b \right|
    \right) = 0.
\end{equation*}

% ======================================

\begin{acknowledgments}
The authors are grateful to A.~E.~Lobanov and A.~V.~Tyukov for useful remarks and fruitful discussions.
\end{acknowledgments}

% ======================================

\end{document}